\documentclass{article}
\usepackage{amsmath,amsthm,amssymb}
\usepackage[dvipdfmx]{graphicx}
\usepackage{pifont}
\usepackage{wrapfig}
\usepackage{subfigure}

\newtheoremstyle{theorem}
  {10pt}		  
  {10pt}  
  {\sl}  
  {\parindent}     
  {\bf}  
  {. }    
  { }    
  {}     
\theoremstyle{theorem}

\newtheoremstyle{defi}
  {10pt}		  
  {10pt}  
  {\rm}  
  {\parindent}     
  {\bf}  
  {. }    
  { }    
  {}     
\theoremstyle{defi}

\begin{document}

\title{\Large{A SEPAK TAKRAW - BASED MOLECULAR MODEL FOR C$_{60}$
- A MATHEMATICAL STUDY OF A 60-ATOM MOLECULE -}}

\author{Yutaka Nishiyama\\
Department of Business Information,\\
Faculty of Information Management,\\ 
Osaka University of Economics,\\
2, Osumi Higashiyodogawa Osaka, 533-8533, Japan\\
nishiyama@osaka-ue.ac.jp\\[2pt]}

\maketitle

\noindent
\textbf{Abstract:} We mathematically investigate four molecular models of buckminsterfullerene (C$_{60}$) discussing the strengths and weaknesses of each, and a new orthorhombic 20-dodecahedron model is proposed to replace the traditional truncated icosahedron model. This representation as a sepak takraw ball rather than a soccer ball allows for capturing the positional relations and electron orbits of the 60 atoms comprising the full molecule.



\section*{1. A new molecular model for a 60-atom molecule}

Spherical molecules composed of 60 atoms such as buckminsterfullerene (C$_{60}$) are often described using a soccer ball truncated icosahedron. Here, I propose a new model for investigating such molecules, one based on an orthorhombic 20-dodecahedron like that shown in Fig. 1. In the figure, black dots represent the 60 atoms, which are situated at vertices of the 20-dodecahedron. Red lines connect individual atoms, and follow the six lines on the ball used in the Southeast Asian sport sepak takraw. We consider these as electron orbits.

I am a mathematician, not a chemist, so what I propose may in the end be nonsense, but I hope that it will be useful as an alternate perspective.

\begin{figure}[htbp]
\begin{center}
\includegraphics[width=40mm]{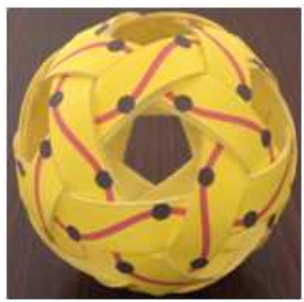}
\end{center}
\caption{A model of a 60-atom molecule. Black dots are atoms, and red lines are electron orbits.}
\end{figure}

\section*{2.  Buckminsterfullerene (C$_{60}$) }

H.W. Kroto et al. discovered the soccer ball-shaped carbon molecule C$_{60}$ in 1985. It was named ``buckminsterfullerene'' after the famous architect Buckminster Fuller. To describe the shape of this molecule, the November 1985 issue of \textit{Nature} [1] included a photo of a soccer ball (Fig. 2). A soccer ball comprises 12 pentagons (black) and 20 hexagons (white), for a total of 32 faces and 60 vertices.

\begin{figure}[htbp]
\begin{center}
\includegraphics[width=120mm]{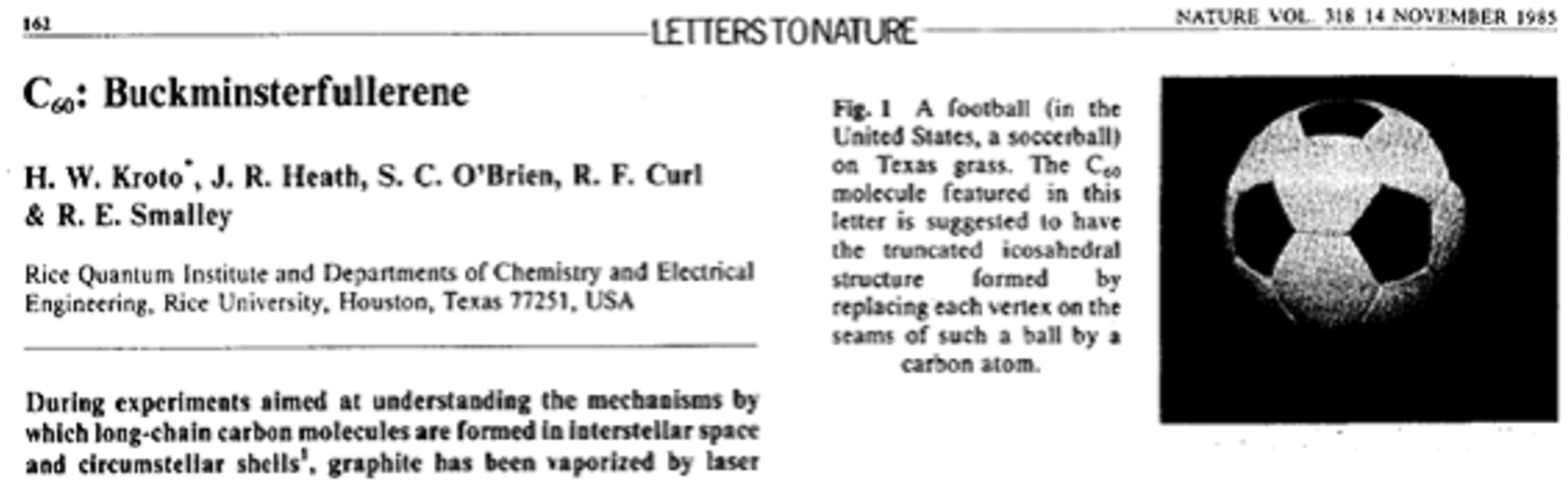}
\end{center}
\caption{A photo of a soccer ball from the Nov 1985 issue of \textit{Nature} [1]}
\end{figure}

The materials shown below explain why a soccer ball was used to describe this molecule's form. Figure 3 is a photo from the New York Times [2] that shows the balls that FIFA used for World Cup matches from 1930 to 2010.

Soccer balls were originally 12 or 18 hexagonal leather panels sewn together. From the 1970 World Cup in Mexico until the 2002 match between Korea and Japan the model instead became a truncated icosahedron formed from 32 panels. This form did not change for approximately 30 years, so the truncated icosahedron became the default image of a soccer ball. The right side of Fig. 3 shows the Telstar ball that was introduced at the 1970 World Cup, and the 1985 \textit{Nature} photo appeared right in the middle of that ball's 30-year reign.

\begin{figure}[htbp]
\begin{center}
\includegraphics[width=100mm]{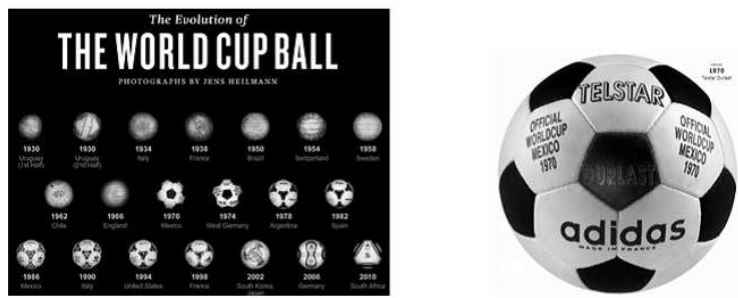}
\end{center}
\caption{Official FIFA soccer balls, 1930-2010 [2]}
\end{figure}

\section*{3. Four current models for representing C$_{60}$ }

In a previous paper [3] I presented four patterns used as a model for the C$_{60}$ molecule. Each is based on a truncated icosahedron, as detailed below.

Referencing Wikipedia pages for ``buckminsterfullerene,'' while there are some differences in the molecular model used for the C$_{60}$ molecule, they generally fall into the following three types (Fig. 4):

\bigskip

(a) Single and double bonds are differentiated, all five-membered rings are single bonds, and six-membered rings have alternating single and double bonds as in a benzene ring.

(b) Single and double bonds are differentiated, but double-bond sites overlap five- and six-membered rings, making placement irregular.

(c) Single and double bonds are not differentiated, and only the positional relation of the sixty atoms is shown.

\bigskip

Examining use of these models by language, the English, Polish, and Chinese Wikipedia pages use (a), the Spanish, Japanese, Persian, Indonesian, Dutch, and Swedish pages use (b), and the French, Danish, Korean, and Italian use (c).

\begin{figure}[htbp]
\begin{center}
\includegraphics[width=110mm]{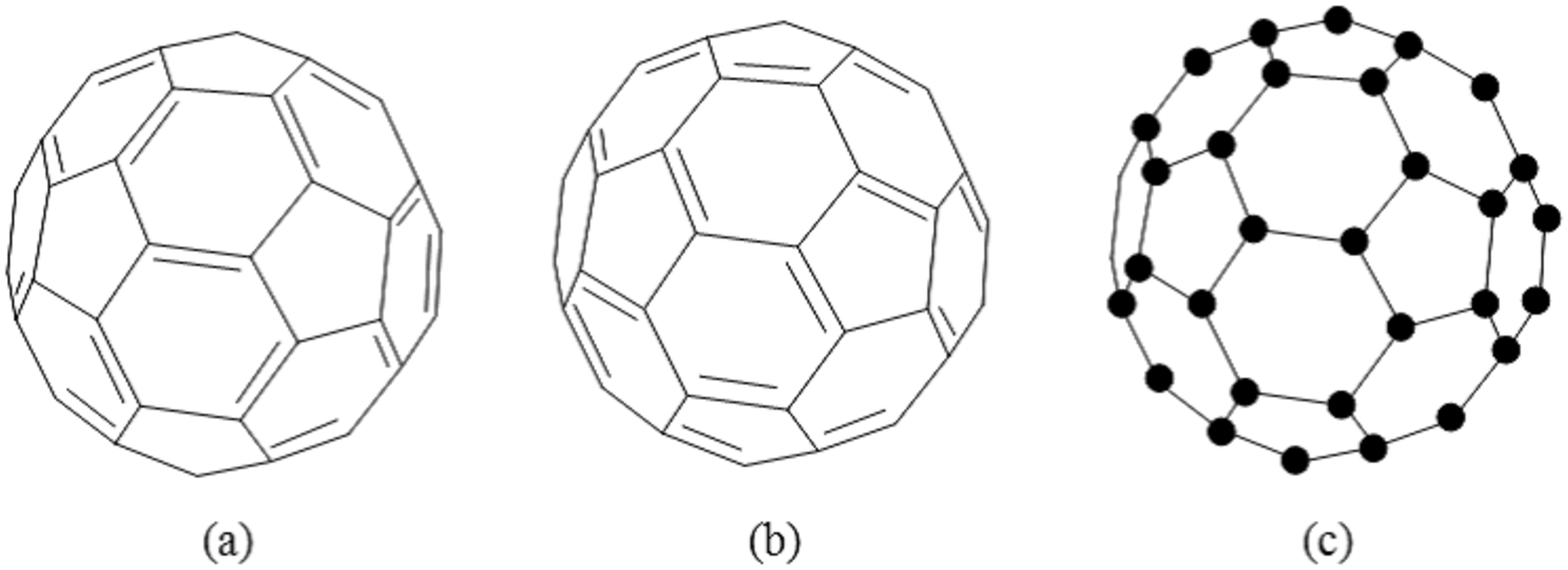}
\end{center}
\caption{Three representations of buckminsterfullerene (C$_{60}$) from Wikipedia [4]}
\end{figure}

In 1991, J.M. Hawkins et al. performed the first X-ray structural analysis of C$_{60}$ [5], a difficult task because each molecule in a C$_{60}$ crystal rapidly spins at room temperature. To do this, they synthesized a C$_{60}$ adduct C$_{60}$(OsO$_{4}$)(4-tert-butylpyridine)$_{2}$ through reaction of C$_{60}$ and osmium tetroxide OsO$_{4}$. This adduct halts the rotation of C$_{60}$ molecules in crystals, sufficiently calming them to allow X-ray structural analysis. Doing so confirmed the position of all carbon atoms in the C$_{60}$ molecule and showed two carbon-carbon distances; distances between adjoining six-membered rings averaged 0.1388(9) nm, and those between five- and six-membered rings averaged 0.1432(5) nm.

This assumes the molecular model of Fig. 4(a), where five-membered rings have only single bonds, but six-membered rings are arranged with alternating single and double bonds. This indicates that three corannulene C$_{20}$H$_{10}$ molecules form a buckminsterfullerene molecule.

The valence of carbon atoms is 4, meaning that there are four bonds extending from each atom. The truncated icosahedron used as the C$_{60}$ model has 60 vertices, 90 edges, and 32 faces, with three edges extending from each vertex. Four bonds therefore extend from each of the three edges, but if each is taken as a single bond then one remains left over. One of the three edges must therefore become a double bond, meaning that each carbon atom will have two single bonds and one double bond. Single bonds have one free electron, while double bonds have two. Double bonds have a higher electron density than do single bonds, thus shortening the bond distance between atoms.

So long as the conditions of one double bond and two single bonds from each carbon atom are satisfied, we are not limited to the molecular model shown in Fig. 4(a). This gives rise to the model of Fig. 4(b), which is no less correct than that of Fig. 4(a); the placement of the double bonds is arbitrary, and will change from time to time. In a previous paper [3] I used a Rubik's Cube to show that the transition from 4(a) to 4(b) is possible. Since both are possible, we can just use the model of Fig. 4(c), which does not show the double bonds, just the 60 carbon atoms. In this case, sides do not represent single bonds.

\begin{figure}[htbp]
\begin{center}
\includegraphics[width=40mm]{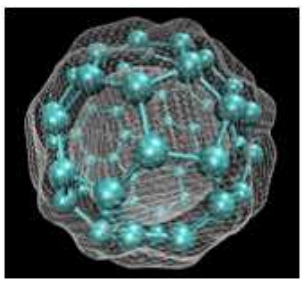}
\end{center}
\caption{Isosurface of ground state electron density for C$_{60}$ [4]}
\end{figure}

The three representations in Fig.4 each have their strengths and weaknesses. Figure 5 shows another image from Wikipedia, that of a C$_{60}$ molecule overlaid with its electron density. Specifically, this is called an isosurface of the ground state electron density, as calculated by the density functional method. In quantum chemistry this is referred to as a cloud of free electrons.

We have so far looked at four current molecular models. So which is correct? In most standard works of chemistry the buckminsterfullerene model is shown as a soccer ball, which gives the impression that the position of each of the 60 atoms in a C$_{60}$ molecule has been clearly identified using X-ray diffraction or electron microscopes. I suspect, however, that such is not the case. What we do know from mass spectrometry is that there are 60 carbon atoms and we know the distance between atoms having single and double bonds.

\section*{4. The metal halide Au$_{144}$X$_{60}$}

In October 2013 the U.S. researcher R.L. Whetten sent me a copy of his paper \textit{Structure  bonding of the gold-subhalide cluster I-Au$_{144}$Cl$_{60}$[z]}, not because we know each other but because in it he had cited my paper related to sepak takraw [6,7]. I am an absolute beginner when it comes to chemistry, so it took me quite some time to decipher his paper, which is related to structural analysis of a molecule called \textit{I}-Au$_{144}$Cl$_{60}$[z].

This molecule comprises 144 gold atoms and 60 chlorine atoms, and with so many atoms this is referred to as a cluster, not a molecule. I also gather that products of a reaction between metallic molecules and the halogen family are called metal halides. Figure 6 shows a diagram of the process by which \textit{I}-Au$_{144}$Cl$_{60}$ is produced [6]. Twelve gold atoms (red) are first formed into Au$_{12}$. This becomes the core around which 30 more gold atoms (orange) are arranged to create Au$_{42}$. A further 12 gold atoms (red) are then added to create Au$_{54}$, then another 60 (blue) to create Au$_{114}$, then another 30 (orange) to create Au$_{144}$. Finally, 60 halogen atoms (green) are added to this metal cluster to create Au$_{144}$X$_{60}$.

\begin{figure}[htbp]
\begin{center}
\includegraphics[width=90mm]{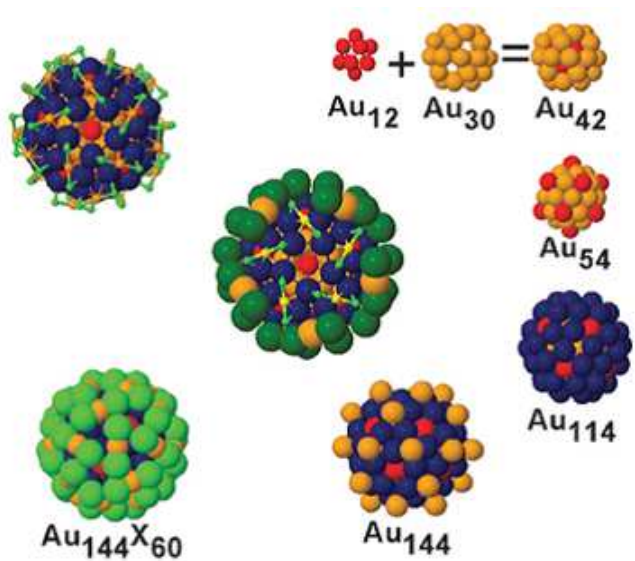}
\end{center}
\caption{Molecular structure of \textit{I}-Au$_{144}$X$_{60}$ [5]}
\end{figure}

The following shows this process as molecular equations:

\bigskip

\hspace{20mm} Au$_{12}$+Au$_{30}$=Au$_{42}$    \qquad \qquad \qquad \qquad     (1)

\hspace{20mm} Au$_{42}$+Au$_{12}$=Au$_{54}$    \qquad \qquad \qquad \qquad     (2)

\hspace{20mm} Au$_{54}$+Au$_{60}$=Au$_{114}$    \qquad \qquad \qquad \quad  \  (3)

\hspace{20mm} Au$_{114}$+Au$_{30}$=Au$_{144}$    \qquad \qquad \qquad \quad     (4)

\hspace{20mm} Au$_{144}$+X$_{60}$=Au$_{144}$X$_{60}$    \qquad \qquad \qquad    (5)

\bigskip

We can think of the arrangement of the 12 gold atoms in Au$_{12}$ as forming an icosahedron. An icosahedron has 20 faces and 12 vertices. Positioning gold atoms on each of these 12 vertices therefore keeps everything in balance. Further adding layers of 30, 12, 60, and 30 gold atoms to the Au$_{12}$ core results in Au$_{144}$ and then combining this with 60 halogen atoms gives Au$_{144}$X$_{60}$. I don't really understand the actual chemical reactions, so I would like to focus on the numbers 12, 30, and 60.

Twelve is associated with dodecahedrons and icosahedrons; a dodecahedron has 12 faces and 20 vertices, and an icosahedron has 20 faces and 12 vertices. Thus, there is a duality relation between the two. For a dodecahedron we can arrange gold atoms on each of the 12 faces, and for an icosahedron on each of the 12 vertices.

For the 30 too, we can consider both dodecahedrons and icosahedrons, since both have 30 edges. We can thus place a gold atom on each edge.

So what about 60? As is well known, there are five types of regular polyhedron (Table 1), namely tetrahedrons (four faces), hexahedrons (six faces), octahedrons (eight faces), dodecahedrons (12 faces), and icosahedrons (20 faces). As the number of faces increases, the polyhedron approaches a sphere. A dodecahedron is denoted by (5, 3), indicating that vertices are formed by three pentagons. Similarly an icosahedron is denoted by (3, 5), indicating five equilateral triangles.

\small

\begin{center}
\begin{tabular}{c|c|c|c|c|c|c}
  & Name & Symbol & Faces (F) & Vertices (V) & Edges (E) & F+V-E \\ \hline
(1)  & Tetrahedron & (3, 3) & 4 & 4 & 6 & 2 \\
(2)  & Hexahedron & (4, 3) & 6 & 8 & 12 & 2 \\
(3)  & Octahedron & (3, 4) & 8 & 6 & 12 & 2 \\
(4)  & Dodecahedron & (5, 3) & 12 & 20 & 30 & 2 \\
(5)  & Icosahedron & (3, 5) & 20 & 12 & 30 & 2 \\ \hline

\end{tabular}
\end{center}

\normalsize

\vspace{2mm}

\begin{center}
Table 1. Regular polyhedrons (Platonic solids)
\end{center}

Regular polyhedrons are defined as having only faces formed by a single type of regular polygon, but Archimedean solids loosen this definition by allowing multiple polygon types. For example, a truncated icosahedron is an Archimedean solid formed by cutting off the vertices of an icosahedron. It has 12 pentagons and 20 hexagons, for a total of 32 faces, 60 vertices, and 90 edges.

There are 13 Archimedean solids. We are interested in a model for molecules like C$_{60}$ with 60 atoms, so focusing on Archimedean solids with 60 vertices we can consider four types: truncated dodecahedrons, truncated icosahedrons, rhombicosidodecahedrons, and snub dodecahedrons. Of these the truncated dodecahedron does not have evenly distributed vertices, so setting that aside we have three solids of interest (Table 2). Each of these includes pentagons.

\small

\begin{center}
\begin{tabular}{c|c|c|c|c|c|c|c|c|c|c}
  & Name & Symbol & Vertices & Edges & Faces  \\ & R/a & S/a$^2$ & V/a$^3$ & V/SR & V/V$_0$ \\ \hline
(1)  & Truncated icosahedron & [5, 6, 6] & 60 & 90 & 32 \\ & 2.478 & 72.607 & 55.288 & 0.3073 & 0.867 \\
(2)  & Rhombicosidodecahedron & [3, 4, 5, 4] & 60 & 120 & 62 \\ & 2.233 & 59.306 & 41.615 & 0.3143 & 0.892 \\
(3)  & Snub dodecahedron & [3, 3, 3, 3, 5] & 60 & 150 & 92 \\ & 2.156 & 55.287 & 37.617 & 0.3156 & 0.896 \\ \hline

\end{tabular}
\end{center}

\normalsize

\vspace{2mm}

\begin{center}
Table 2. Archimedean solids with 60 vertices

\vspace{2mm}

\scriptsize{
(R: Radius of circumscribed sphere, a: Edge length, S: Face surface, V: Volume, 
V/SR: 0.3333 for circumscribed sphere, V/V$_0$: Volume ratio of circumscribed sphere)}

\end{center}

In a clockwise direction around each vertex of a truncated icosahedron is a pentagon, a hexagon, and another hexagon, so it is designated by [5, 6, 6]. Similarly, around each vertex of a rhombicosidodecahedron is an equilateral triangleCa tetrahedronCa pentagon, and a tetrahedron, and so is designated by [3, 4, 5, 4]. A snub dodecahedron is [3, 3, 3, 3, 5].

We want to focus particularly on the truncated icosahedron and the rhombicosidodecahedron. The ratios V/SR and V/V$_0$ are used as units for measuring the circumscribed sphere of an Archimedean solid. When V/SR is 0.3333 or V/V$_0$ is 1, the object is a sphere. Therefore, a rhombicosidodecahedron is closer to a sphere than is a truncated icosahedron.

Of the molecular equations given above, I paid particular attention to Eqs. (3) and (5). 

\bigskip
\hspace{20mm} Au$_{54}$+Au$_{60}$=Au$_{114}$    \qquad \qquad \qquad \quad  \  (3)

\hspace{20mm} Au$_{144}$+X$_{60}$=Au$_{144}$X$_{60}$    \qquad \qquad \qquad    (5)

\bigskip

In (3) there are 60 gold atoms, and in (5) there are an additional 60 halogen atoms. I was particularly interested in how the 60 gold atoms would be arranged in (3).

There is a public domain software package called Jmol that can create three-dimensional images of chemical structures. I used this software to create images of Au$_{144}$X$_{60}$ (Fig. 7(a)) and Au$_{114}$ with its interior Au54 removed (Fig. 7(b)), which leaves the latter with 60 gold atoms.

\begin{figure}[htbp]
\begin{center}
\includegraphics[width=90mm]{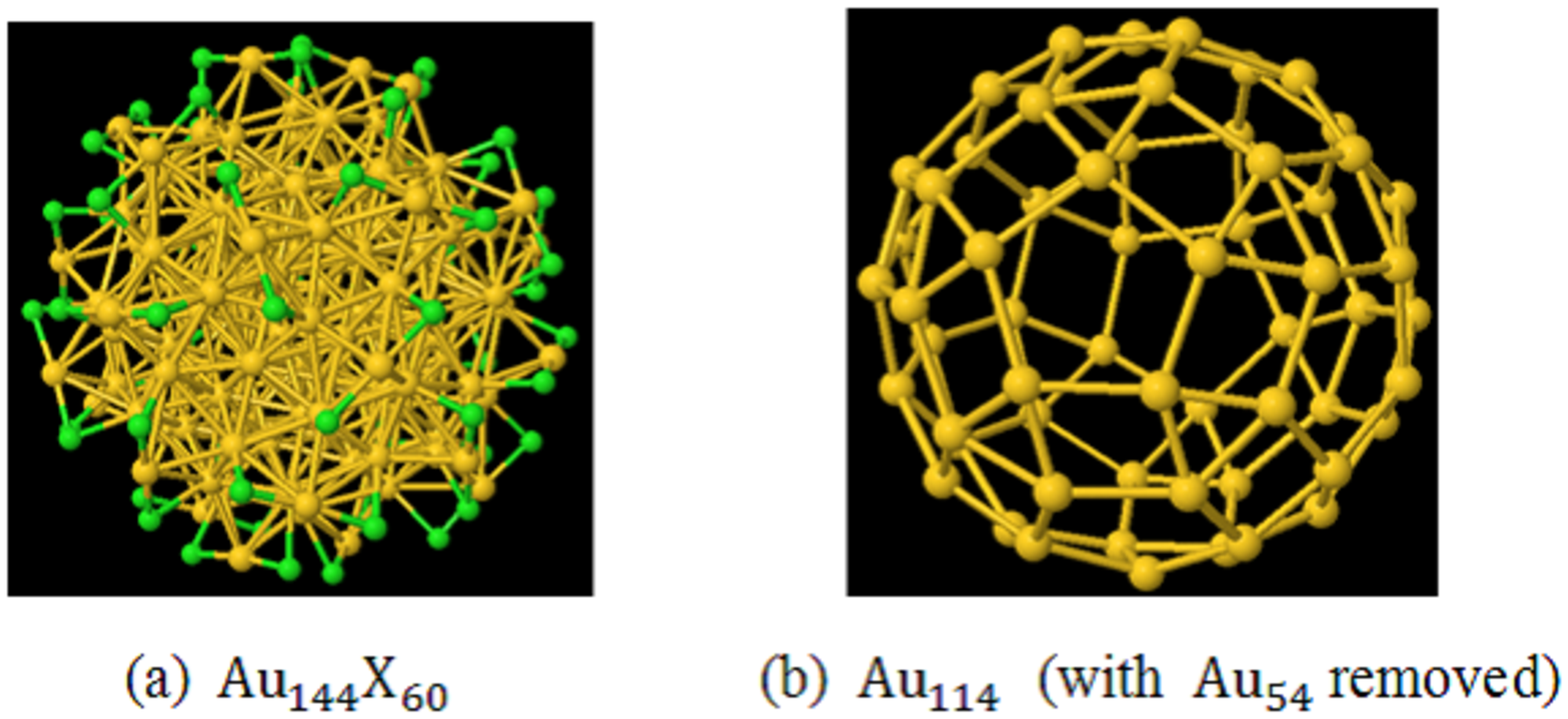}
\end{center}
\caption{Images from Jmol}
\end{figure}

Closely observing the 60 gold atoms in Fig. 7(b), we can see that their arrangement is subtly different from that of a C$_{60}$ molecule. C$_{60}$ used a truncated icosahedron as its model, but these 60 atoms are arranged as a rhombicosidodecahedron. In other words it is formed not by pentagons and hexagons, but by equilateral triangles, tetrahedrons, and pentagons. Until noticing this I had thought that the only molecular model for arranging 60 atoms was a truncated icosahedron, so this was quite a surprise.

\section*{5. A new molecular model}

As discussed above, the standard molecular model for C$_{60}$ is a soccer ball. Another ball that has a similar truncated icosahedron is the ball used in sepak takraw. A truncated icosahedron has 12 pentagons and 20 hexagons, for a total of 32 faces, and in a soccer ball these faces are radially inflated into a sphere (Fig. 8(a)). A sepak takraw ball in contrast has six woven bands that trace along six great circles of an icosahedron, the resulting spaces between bands forming pentagons. Figure 8(b) shows a competition sepak takraw ball, and Fig. 8(c) shows one for mathematical puzzles woven out of polypropylene bands. I will use the ball in Fig. 8(c) to compare the truncated icosahedron and the rhombicosidodecahedron (Fig. 9).

\begin{figure}[htbp]
\begin{center}
\includegraphics[width=100mm]{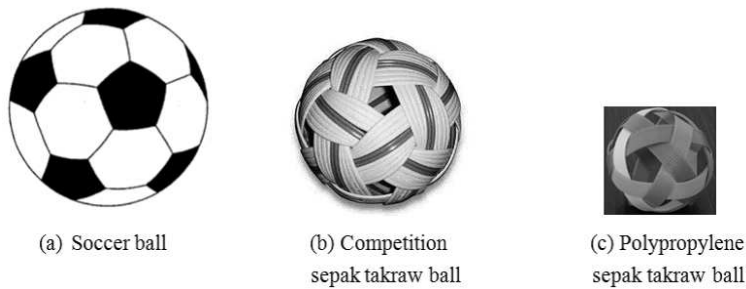}
\end{center}
\caption{Soccer ball and sepak takraw balls}
\end{figure}

\begin{figure}[htbp]
\begin{center}
\includegraphics[width=110mm]{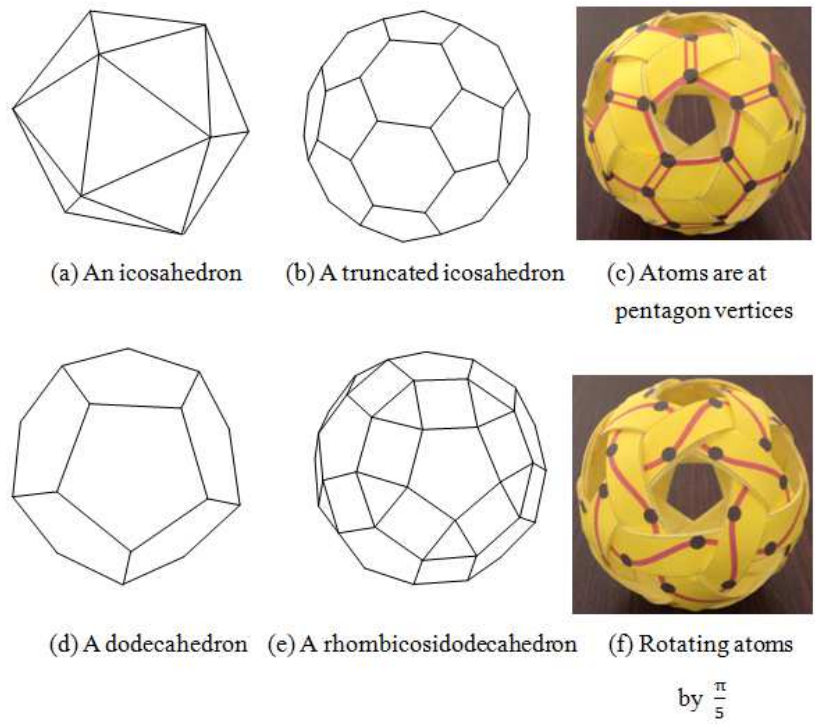}
\end{center}
\caption{Truncated icosahedron and rhombicosidodecahedron}
\end{figure}

First, the icosahedron in Fig. 9(a) and the dodecahedron in Fig. 9(d) are in a mathematical duality. Lopping off the vertices of an icosahedron results in the truncated icosahedron of Fig. 9(b), a molecular model of C$_{60}$. Similarly trimming the vertices of a dodecahedron gives the rhombicosidodecahedron of Fig. 9(e), a molecular model of Au$_{60}$ (Fig. 7(b)).

Soccer balls and sepak takraw balls can be created from truncated icosahedrons. As mentioned above, a sepak takraw ball can be formed from six bands, with the resulting empty spaces forming pentagons. Figure 9(c) shows a sepak takraw model with atoms (black dots) at the vertices of these empty spaces and atoms connected by one or two bonds (red lines). Single lines indicate single bonds, and double bonds are shown by two lines over places where two bands cross. This arrangement of atoms and bonds gives exactly the molecular model for C$_{60}$ shown in Fig. 4(a). In terms of that figure, the sepak takraw ball and the soccer ball are the same.

I discussed with Prof. Whetten the molecular model for the metal halide Au$_{144}$X$_{60}$ described above. He said that he was unable to obtain high-quality polypropylene bands in the U.S., so I sent him some from Japan. This allowed us to discuss how features of the sepak takraw ball can be utilized.

As part of our trial-and-error investigations, he proposed a model where the atoms (black points) of Fig. 9(c) are rotated to the positions shown in Fig. 9(f). In that rearrangement the atoms on pentagon vertices are rotated clockwise around the pentagon center by $\pi/5$ and relocated on pentagon edges. I tried connecting the moved atoms using red lines. There are 60 atoms in all, and six bands. When they are connected in this way the 60 atoms are divided among the six bands, so that $60 / 6 = 10$ atoms are placed on each band.

\begin{figure}[htbp]
\begin{center}
\includegraphics[width=120mm]{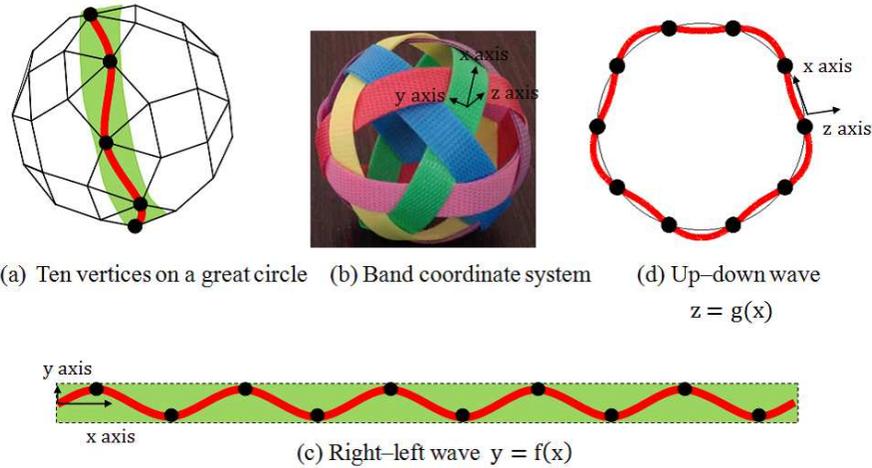}
\end{center}
\caption{Atom positions and electron orbits in a rhombicosidodecahedron model}
\end{figure}

Let's take a closer look at Fig. 9(f). The six bands in a sepak takraw ball are woven in an up-down-up-down pattern. In places where three bands cross, they are tightly affixed in a triple relation. Figure 10(a) shows one of the six bands (green) following a great circle of the rhombicosidodecahedron, with a total of 10 atoms (black dots) on vertices, connected by a closed sine wave curve (red). 

We construct on the bands a three-dimensional $xyz$ coordinate system like that shown in Fig. 10(b) to allow representation of electron orbits using equations. Taking an arbitrary point on a band as the origin, we take the band direction as the $x$-axis, the left-right direction as the $y$-axis, and the direction from the center of the ball to the outside as the $z$-axis. We thus specify $(x, y, z)$ coordinates using a right-handed coordinate system, with the $x$-, $y$-, and $z$-axes respectively in the direction of the thumb, index finger, and middle finger of a right hand.

Letting electron orbits (red) move along the sine wave $y=f(x)$ to the right and left with respect to the direction of the band (up and down in the figure), we can consider a composition with an up down sine wave $z=g(x)$ (Figs. 10(c) and (d)).

Taking the radius of a sepak takraw ball as $r$, and the circumference of a great circle around the ball as $L$, then that circumference is $L=2\pi r$. We can represent two sine waves on this great circle band, both with cycle 5, by the following equations:

$$ y=f(x)=a \sin(5x-\alpha)       \qquad \qquad \qquad \qquad     (6) $$
$$ z=g(x)=b \sin(5x-\beta)         \qquad \qquad \qquad \qquad     (7) $$
   $$ - \infty <x<+ \infty  \qquad \qquad \qquad \qquad $$

\noindent
Here, $a$ and $b$ are wave amplitudes and $\alpha$ and $\beta$ are phase shifts.

On this belt are 10 atoms and the free atoms associated with them. The six electron orbits follow the left-right, up-down sine waves with differing phases, so the orbits will not overlap. You can imagine this as the up-down-up-down interweaving of the bands in a sepak takraw ball.

Note that a sepak takraw ball is woven from six bands, but its size is fixed without the use of adhesives. Taking the width of a band as $d$, the ball circumference as $L$, the ball radius as $r$, and its diameter as $D$, then $L, D$, and $d$ have the relationship given by the following equations [7]:

$$ L=10 \sqrt{3} d     \qquad \qquad \qquad \qquad \qquad \qquad \qquad  (8)$$
$$ D=2r=\frac{L}{\pi} = \frac{10 \sqrt{3}}{\pi} d \approx 5.51d    \qquad \qquad \quad \; \; (9)$$

From this, bands with width 15 mm result in a sepak takraw ball with a diameter of approximately 83 mm.

This made me wonder if the left-right width of the electron orbits might not determine the size of the molecule, just as the band width determines the size (diameter) of a sepak takraw ball. In other words, if we take the electron orbits given by Eqs. (6) and (7) as the band width $d$ in Eqs. (8) and (9), might this determine a molecule size $D$?

Molecules are formed from atoms. Van der Waals'forces act between those atoms, making them attractive when too far apart and repulsive when too close together. The molecule is stable when there is a certain distance between its atoms. There are covalent, ionic, and metallic chemical bonds, and those in a C$_{60}$ molecule are covalent, and the valence of its carbon atoms is 4. The gold cluster Au$_{60}$ has metallic bonds, and its gold atoms have valence values of 1 or 3. It would thus be going too far to consider a 60-atom molecular model taking C$_{60}$ and Au$_{60}$ as equivalent, but there are some advantages to changing the truncated icosahedron to a rhombicosidodecahedron:

\bigskip

(1) The positions of the atoms better approximate a sphere.

(2) There will be four bonds connecting each carbon atom, removing the problem of double bonds.

\bigskip

Furthermore, we obtain a sense of unity of the molecule overall, not unlike the six bands in a sepak takraw ball. A soccer ball is created by sewing together 12 pentagon (black) and 20 hexagon (white) leather panels. While there is a relationship between adjacent panels, this does not give a sense of unity for the whole. In contrast, a sepak takraw ball is formed from six interwoven bands that trace out great circles, which gives it a much more unified feel. Considering not only the relation between five atoms (five-membered rings) and six atoms (six-membered rings), but the bundles of all 60 atoms into six electron orbits, the latter feels the better suited.

\bigskip

In the above we have considered a model for a 60-atom molecule, but as previously stated I am a mathematician, with almost no knowledge of chemistry. While my observations may be nothing but fantasy, I hope they will be of interest to some.


\begin{thebibliography}{99}

\bibitem{kro} H. W. Kroto, J. R. Heath, S. C. O'Brien, R. F. Curl \& R. E. Smalley, C$_{60}$: Buckminsterfullerene, \textit{Nature}, \textbf{318}, (1985), 162-164. DOI: 10.1038/318162a0

\bibitem{fifa} The Evolution of the World Cup Ball, New York Times, 6th June 2014,

http://www.nytimes.com/interactive/2010/06/06/magazine/20100606-world-cup-balls.html (accessed March 2015).

\bibitem{nishi1} Y. Nishiyama, A C$_{60}$ Fullerene Model and Sepak Takraw Balls, \textit{Int. J. Pure Appl. Math.}, \textbf{94} (5), (2014), 669-688.

doi: http://dx.doi.org/10.12732/ijpam.v94i5.4


\bibitem{wiki} Wikipedia contributors, ``Buckminsterfullerene,'' Wikipedia, The Free Encyclopedia, http://en.wikipedia.org/wiki/Buckminsterfullerene (accessed March 2015).


\bibitem{haw} J. M. Hawkins, et al., Crystal Structure of Osmylated C$_{60}$: Confirmation of the Soccer Ball Framework, \textit{Science}, \textbf{252}, (1991), 312-313. DOI: 10.1126/science.252.5003.312


\bibitem{tla} A. Tlahuice-Flores, R. L. Whetten, et al., Structure \& bonding of the gold-subhalide cluster \textit{I}-${\rm Au}_{144}{\rm Cl}_{60}$$^{[z]}$, \textit{Phys.  Chem. Chem. Phys.}, \textbf{15}, (2013). DOI: 10.1039/C3CP53902D


\bibitem{nishi2} Y. Nishiyama, The Sepak Takraw Ball Puzzle, \textit{Int. J. Pure Appl. Math.}, \textbf{79}(2), (2012), 281-291.


\end{thebibliography}
\end{document}